\documentclass[12pts, epsf,aps,prb]{revtex4}
\usepackage{graphicx} 
\begin{document}
\title{Field induced metamagnetic transitions in quasi-one-dimensional Ising spin chain CoV$_{2}$O$_{6}$}

\author{M. Nandi and P. Mandal}
\email{prabhat.mandal@saha.ac.in}
\affiliation{Saha Institute of Nuclear Physics, 1/AF Bidhannagar, Calcutta 700 064, India}
\date{\today}

\begin{abstract}
We have investigated the temperature and magnetic field dependence of magnetization ($M$), specific heat ($C_p$), and  relative sample length change ($\Delta L/L_0$) for understanding the field-induced metamagnetic transitions in quasi-one-dimensional monoclinic ($\alpha$ phase) and triclinic ($\gamma$ phase) CoV$_2$O$_6$ spin chains. With the application of external magnetic field, a sharp peak emerges in $C_p$($T$) below $T_N$$=$15 K in $\alpha$ phase and a hump-like feature appears above $T_N$$=$6.5 K in $\gamma$ phase due to the field-induced ferrimagnetic and ferromagnetic transitions, respectively. Strong field dependence of linear thermal expansion and large positive magnetostriction have been observed in $\alpha$ phase. Though magnetization data indicate strong spin-orbit coupling for both phases, other measurements show that this effect is very weak for the $\gamma$ phase.
\end{abstract}
\newpage
\maketitle
\section{Introduction}
Cobalt based low-dimensional systems have drawn considerable attention in condensed matter physics due to their complex magnetic behavior. Such compounds having spin-chain structure exhibit strong magnetic anisotropy, field-induced metamagnetic transitions, 1/3 magnetization plateau, etc. For example, Ca$_3$Co$_2$O$_6$, which is constituted of Co$^{3+}$ ion chains, exhibits cascade of magnetization plateaux.\cite{maignan,maignan1} Sometimes the plateau attains a value which is just 1/3 of the saturation magnetization such as in cobaltate dihydrate (CoCl$_2$:2H$_2$O).\cite{koba} Particularly, quasi one dimensional Ising spin chains display various fascinating magnetic phenomena like quantum criticality, order-disorder transition, etc. Ising spin chain CoNb$_2$O$_6$ is quite famous for the experimental evidence of quantum phase transition in transverse field.\cite{cold} For this system, an  E$_8$ symmetry  has been experimentally evidenced near the quantum critical point.  Another well known Ising chain BaCo$_2$V$_2$O$_8$  exhibits peculiar order to disorder transition  in presence of magnetic field.\cite{he}\\

In recent years, quasi-one-dimensional Ising spin chain CoV$_{2}$O$_{6}$ has attracted immense interest because
of the diverse magnetic transitions, large single ion anisotropy, strong spin-lattice coupling, etc. Depending on the local environment, CoV$_{2}$O$_{6}$ displays  two different types of crystalline phases.\cite{lene,kim,holl} The high-temperature $\alpha$ phase crystallizes in a brannerite-like monoclinic structure,\cite{zhe,lene,kim,holl,lener,ksingh,mark,markk,nandi} while the low-temperature $\gamma$ phase exhibits a triclinic structure.\cite{zh,lene,kim,kimb,holl} In both the phases, the edge-sharing CoO$_6$ octahedra form magnetic chains along the $b$ axis. These chains are separated by zigzag chains of VO$_5$ pyramids and VO$_6$ octahedra for the $\alpha$ and $\gamma$ phase, respectively. The easy axis is perpendicular to chain direction in $\alpha$ phase \cite{zhe,lener} while it is pointing along the chain direction in $\gamma$ phase.\cite{lenert,lenertz}  The magnetic properties of CoV$_{2}$O$_{6}$ are solely determined by the divalent Co$^{2+}$ ions because the V ions are in a nonmagnetic V$^{5+}$ state. The Co$^{2+}$ ions within the same chain are strongly coupled via ferromagnetic (FM) interaction while the interchain interaction is weak and antiferromagnetic (AFM) in nature. Both $\alpha$ and $\gamma$ phases exhibit AFM to paramagnetic (PM) transition around 14 K and 6.5 K, respectively.\cite{zhe,lene,kimb}In this work, we present our results on  magnetization, heat capacity and thermal expansion study for both $\alpha$-CoV$_{2}$O$_{6}$ and $\gamma$-CoV$_{2}$O$_{6}$ compounds. Several important differences between field-induced magnetic states in these two systems are also discussed.\\

\section{Experimental details}

Polycrystalline  $\alpha$-CoV$_{2}$O$_{6}$ and $\gamma$-CoV$_{2}$O$_{6}$ samples were prepared by standard solid-state reaction method using the mixture of stoichiometric quantities of high purity V$_2$O$_5$ and cobalt acetate tetrahydrate or cobalt oxalate dihydrate. For $\alpha$-CoV$_{2}$O$_{6}$,  the mixture was heated in air for 16 h at 650$^{\circ}$C and then at 725$^{\circ}$C for 48 h. After the heat treatment, the material was quenched in liquid nitrogen  to obtain single phase $\alpha$-CoV$_{2}$O$_{6}$. For $\gamma$-CoV$_{2}$O$_{6}$, the mixture was heated in air at 620$^{\circ}$C for 45 h and then cooled to room temperature at a rate of 2$^{\circ}$C/min. The resulting powder was ground properly and pressed into pellets. Finally, the pellets were heated again at 620$^{\circ}$C  for 45 h and cooled at a rate of 2$^{\circ}$C/min. Phase purity of these compounds was checked by powder x-ray diffraction (XRD) method with CuK$_{\alpha}$ radiation in a  Rigaku TTRAX II  diffractometer. No trace of impurity phase was detected within the resolution of XRD. All the peaks in XRD were assigned to a monoclinic structure of space group $C$2/$m$ for $\alpha$ phase whereas to a triclinic structure of space group P-1 for $\gamma$ phase using the Rietveld method. The magnetization measurements were done using a SQUID-VSM (Quantum Design). The specific heat measurements were done using a physical property measurement system (Quantum Design) by conventional relaxation time method. The thermal expansion measurements were done by capacitive method using a miniature tilted-plates dilatometer.\\

\section{Results and discussion}

The zero-field cool (ZFC) susceptibility ($\chi$) and inverse susceptibility ($\chi^{-1}$) versus temperature curves for both $\alpha$ and $\gamma$ phases are presented in Figure 1. Measurements were performed at 0.1 T for both $\alpha$ phase and $\gamma$ phase samples. For $\alpha$ phase, $\chi$ increases monotonically with decreasing temperature and exhibits a sharp peak around $T_N$$=$14 K due to the transition from PM to AFM state. Similar to $\alpha$ phase, $\chi$($T$) of $\gamma$ phase also exhibits a sharp peak around 6.5 K due to the AFM transition. We observe that $\chi^{-1}$($T$) for both the compounds can be fitted well with the Curie-Weiss law [$\chi$$=N\mu_{eff}^2/3k(T-\theta$)] over a wide range of temperature as shown in Figure 1. From this Curie-Weiss fitting, we have calculated effective paramagnetic moment $\mu_{eff}$$=$5.3 $\mu_B$/Co ion and the Weiss temperature $\theta$$=$-2.4 K  for the $\alpha$ phase while the corresponding values are 5.2 $\mu_B$/Co ion and -7 K  for the $\gamma$ phase. The negative $\theta$ indicates that the predominant magnetic interaction in these systems is antiferromagnetic in nature. The value of effective moment in paramagnetic state for both the compounds is much larger than the expected spin-only moment of high spin Co$^{2+}$ (3.87 $\mu_B$). We would like to mention that these values of $\mu_{eff}$ are comparable with those reported for  polycrystalline and single crystalline samples.\cite{mark,kimb,lenertz,Drees}

Isothermal magnetization curves at low temperature are plotted in Figure 2. The data were collected for field increasing condition at 5 K for the $\alpha$ phase and at 2 K for the $\gamma$ phase. Even though these samples are magnetically random, particularly for $\alpha$ phase two sharp metamagnetic transitions are observed around two critical fields $H_{c1}$ and $H_{c2}$. $\alpha$-CoV$_{2}$O$_{6}$ exhibits field-induced transitions from AFM to ferrimagnetic (FI) state at $H_{c1}$ ($\sim$ 1.5 T) followed by FI to FM state at $H_{c2}$ ($\sim$ 3.3 T) with a 1/3 magnetization plateau when the magnetic  field is applied parallel to the $c$ axis. Similar to $\alpha$-CoV$_{2}$O$_{6}$, $\gamma$-CoV$_{2}$O$_{6}$ also exhibits two successive metamagnetic transitions with 1/3 magnetization plateau along easy axis. However, the anomalies around two critical fields for the $\gamma$-CoV$_{2}$O$_{6}$ are not clearly visible in $M$($H$) curve even at 2 K.  To track the critical fields, we have calculated the derivative of $M$($H$) which exhibits two peaks around two critical fields. The critical fields $H_{c1}$ and $H_{c2}$ determined in this way are about 0.4 T and 0.6 T, respectively. These values are close to earlier reports.\cite{lene,kimb} Two field-induced transitions are clearly observed even for the magnetically random  $\alpha$-phase sample while these two transitions are not clearly separated in the $M$($H$) curve of magnetically random $\gamma$-phase sample. Actually, the values of critical fields as well as the difference between two critical fields for the $\gamma$ phase are very small as compared to that for the $\alpha$ phase. For this reason, the anomalies around two critical fields are not distinguishable in magnetically random $\gamma$-phase sample. Though the chemical formula is same for both phases, the structures are quite different. The strength of magnetic interaction is mainly determined by the interatomic distances. In $\gamma$-phase, the interchain distance is larger compared to $\alpha$ phase.\cite{lene} Due to the larger interachain distance, the  antiferromagnetic interaction is weaker in $\gamma$-phase as compared to $\alpha$-phase which is reflected in their ground state ordering temperature.   The  antiferromagnetic ordering temperature  for  the $\gamma$-phase  is just half of that for the $\alpha$-phase.  As the interchain coupling is much weaker in $\gamma$-phase, a smaller field is required to align the chains ferromagnetically.  For this reason, $\gamma$-phase has smaller critical fields as compared to that for the $\alpha$-phase.

Specific heat ($C_p$) versus temperature curves in the vicinity of $T_N$ are shown in Figures 3 (a) and (b) for some selected magnetic fields for  $\alpha$-CoV$_{2}$O$_{6}$ and $\gamma$-CoV$_{2}$O$_{6}$, respectively. At zero field, a $\lambda$-like peak has been observed around $T_N$. For the $\alpha$ phase, when the external magnetic field is applied, the peak at $T_N$ gradually suppresses, broadens and shifts towards lower temperature. Apart from this,  an interesting behavior with application of magnetic field is observed. For example, a new peak appears around 11 K for applied field of 2.5 T due to the field-induced FI transition. This field-induced peak remains clearly visible even at field as high as 8 T. To determine the exact positions of the peaks, we have deconvoluted the peaks.  A pair of Gaussian functions plus a suitable background was fitted to the data. It is observed that the peak positions shift slowly with field. Also, the field-induced peak is found to be quite sharp, possibly due to the first-order nature of the transition. Unlike $\alpha$-CoV$_{2}$O$_{6}$, a clear hump-like feature appears well above $T_N$ in $C_p$ data of $\gamma$-CoV$_{2}$O$_{6}$ due to the emergence of a FM phase when the applied magnetic field exceeds a critical value.  For example, this hump-like feature occurs around 9.5 K ($\sim$$T_F$) at 2 T and shifts slowly towards higher temperature with increase of magnetic field. Closer inspection reveals that another very weak anomaly appears around 5 K ($\sim$$T_{FI}$) due to the FI phase.  Though both these two features are observable at  1 T, they are more clearly visible at 2 T and above [inset of Figure 3(b)]. With the increase of magnetic field strength, the hump-like feature pronounces whereas the peak at $T_N$ and the weak anomaly just below $T_N$ progressively suppress. At 5 T and above,  only the hump-like feature survives. In a magnetically random sample, different ordered magnetic phases coexist. When the applied field is greater than $H_{c1}$ ($H_{c2}$), the system contains a certain fraction of FI (FI and FM) phase. The reported low-temperature neutron diffraction results also support this observation. \cite{lener} For $\alpha$ phase, powder neutron diffraction studies have revealed that both AFM and FI phases coexist with percentage of 54 and 46, respectively at 2.5 T. At fields above $H_{c2}$, however, AFM, FI and FM phases coexist and with further increase of field strength AFM and FI fractions decrease significantly due to the increase of FM phase.

In order to investigate the spin-orbit coupling and field-induced transitions in more details, we have also studied the linear expansion of sample length [$\Delta L/L$] as  functions of $T$ and $H$. For $\alpha$-CoV$_{2}$O$_{6}$, the temperature dependence of $\Delta L$($T$)/$L_{4K}$ ($=$[$L$($T$)-$L_{4K}$]/$L_{4K}$) up to 35 K for some selected fields has been plotted in Figure 4 (a), where $L_{4K}$ is the length of the sample at $T$$=$4 K in an applied field $H$.  At zero field, $\Delta L/L_{4K}$ shows a sharp anomaly around $T_N$.  At 2.5 T, $\Delta L$($T$)/$L_{4K}$ exhibits a step-like  decrease below the antiferromagnetic transition temperature. For understanding the field-induced anomalies at and below $T_N$, we have calculated thermal expansion coefficient ($\alpha$) at different fields. The temperature dependence of  $\alpha$ is shown in Figure 4(b). It is clear from the figure that $\alpha$($T$) curves are qualitatively similar to that of $C_p$($T$) curves. At zero field, $\alpha$($T$) exhibits a $\lambda$-like peak around $T_N$. With the application of magnetic field, a very sharp peak appears below $T_N$ which remains visible up to 4 T. Thus, similar to specific heat data, the field-induced transitions are reflected in thermal expansion measurements. For the polycrystalline sample, though the linear thermal expansion gives an average expansion along three crystallographic axis directions, the field-induced features in thermal expansion is observed mainly due to the change of $c$ axis length.  Our results on thermal expansion is consistent with reported data of temperature variation of lattice parameters determined using powder neutron diffraction \cite{mark,markk}.\\

Figure 5 presents a comparison between the magnetostriction, $\Delta L$($H$)/$L_0$$=$[$L$($H$)-$L_0$]/$L_0$, where  $L_0$ is the length of the sample in absence of magnetic field, and magnetization  for $\alpha$-CoV$_2$O$_6$. It is clear from Figure 5 that the magnetostriction is very sensitive to temperature and displays a huge value at low temperature. At 3 K, $\Delta L$/$L_0$ exhibits two sharp anomalies around two critical fields of metamagnetic transitions. Magnetostriction curve also shows strong hysteresis similar to that observed in magnetization data at 3 K. Above $T_N$, these two anomalies disappear in magnetostriction as well as in magnetization curve. We have also measured magnetostriction at temperatures well above $T_N$ and observed no anomaly. For example,  $\Delta L$/$L_0$ at 35 K increases monotonically with temperature like magnetization curve (inset of Figure 5). It is clear from Figure 5 that $\Delta L$($H$)/$L_0$ is large and approximately mimics the nature of $M$($H$) curve for $\alpha$ phase. So it can be concluded that there exists a strong spin-lattice coupling in $\alpha$ phase.
We would like to mention that the linear expansion of sample length  as  functions of both $T$ and $H$  have also been measured for $\gamma$ phase.
In contrast to $\alpha$ phase, a very weak anomaly has been observed in $\Delta L/L_{4K}$ versus $T$ curve around $T_N$ and no other field-induced transition has been found.  The value of  magnetostriction is also very small as compared to $\alpha$ phase.\\

\section{Conclusions}
In conclusion, we have compared and contrasted different physical properties of quasi-one-dimensional antiferromagnetic  monoclinic and triclinic spin chains of CoV$_2$O$_6$. For both the phases, $C_p$($T$) exhibits a sharp $\lambda$-like peak around $T_N$. Though $C_p$ of both phases exhibit strong $T$ and $H$ dependence, appeared features in $C_p$($T$) due to the external magnetic field are different for two phases. When the applied field exceeds a certain critical field, a sharp peak appears below $T_N$ due to the ferrimagnetic-paramagnetic transition for $\alpha$ phase whereas a weak anomaly below $T_N$ and a hump-like feature above $T_N$ appear in $\gamma$ phase due to FI-PM and FM-PM transitions respectively. Similar to $C_p$($T$), thermal expansion coefficient also exhibits emergence of field-induced peak below $T_N$ in $\alpha$-CoV$_2$O$_6$. Huge magnetostriction effect has been found below $T_N$ in $\alpha$ phase. So, it indicates that the spin-lattice coupling in $\alpha$ phase is quite strong. For $\gamma$ phase, linear expansion $\Delta L/L_{4K}$ exhibits a very weak anomaly around $T_N$ and no significant amount of magnetostriction has been observed. So it can be concluded that the spin-lattice coupling is very weak in $\gamma$ phase compared to $\alpha$ phase.

\newpage
\begin{figure}[h]
\includegraphics[width=0.6\textwidth]{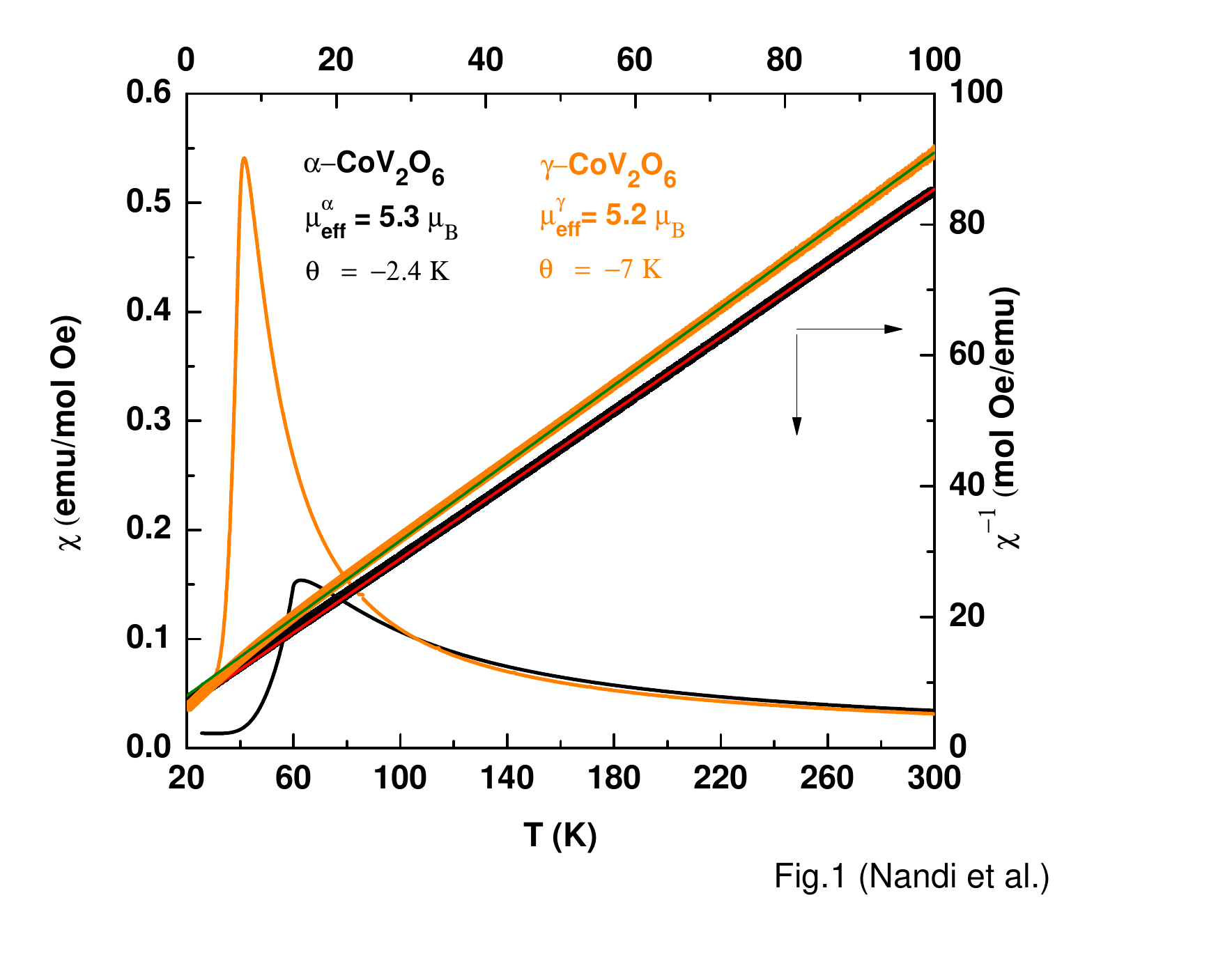}
\caption{Temperature dependence zero-field-cool susceptibility ($\chi$) for $\alpha$-CoV$_{2}$O$_{6}$ and $\gamma$-CoV$_{2}$O$_{6}$ at 0.1 T. The right axis shows the inverse of susceptibility  ($\chi$$^{-1}$)  and the corresponding Curie-Weiss fit (solid line).}\label{fig1}
\end{figure}

\begin{figure}[h]
\includegraphics[width=0.6\textwidth]{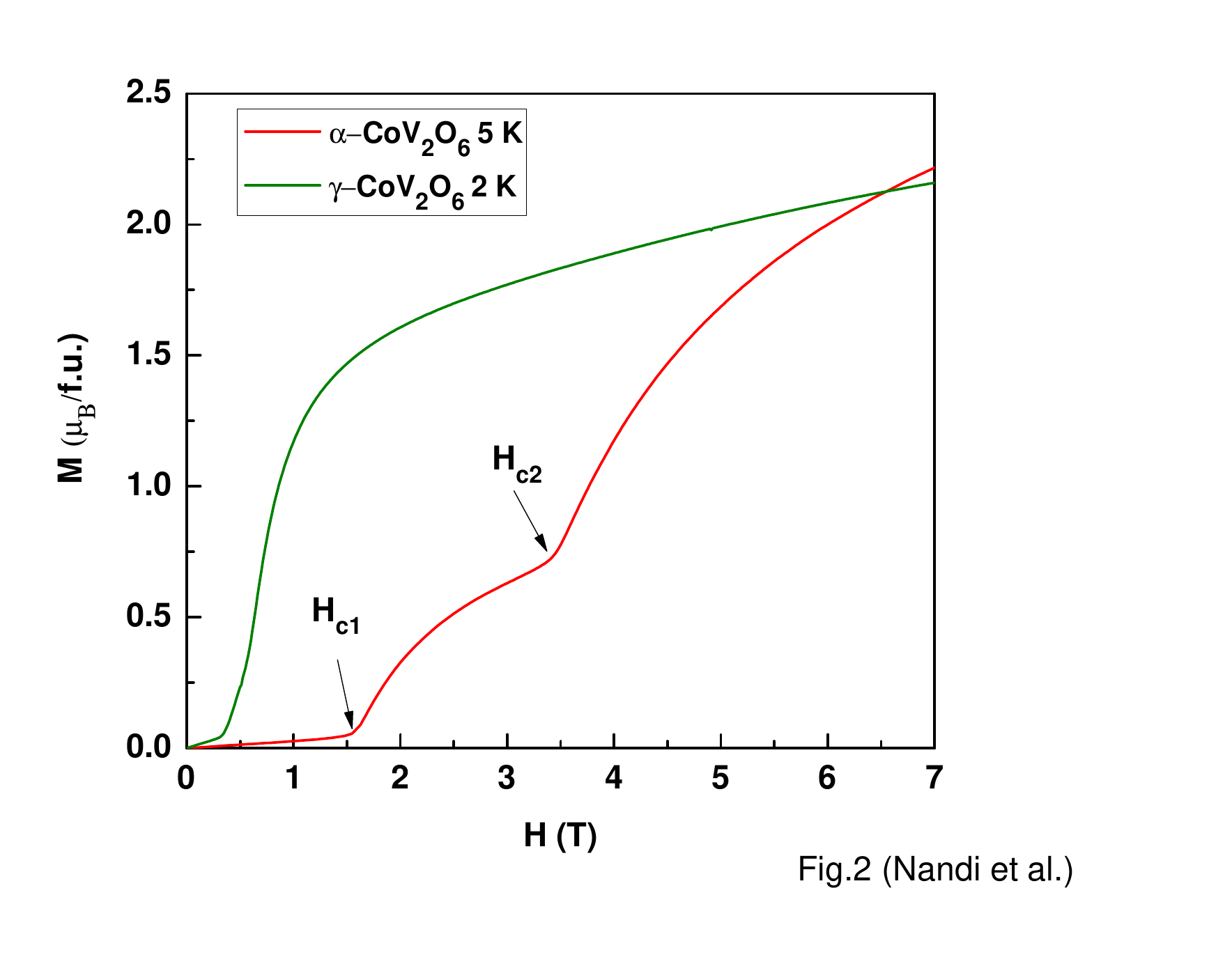}
\caption{Isothermal magnetization for $\alpha$-CoV$_{2}$O$_{6}$ and $\gamma$-CoV$_{2}$O$_{6}$ at 5 K and 2 K respectively.}\label{fig2}
\end{figure}

\begin{figure}[h]
\includegraphics[width=0.6\textwidth]{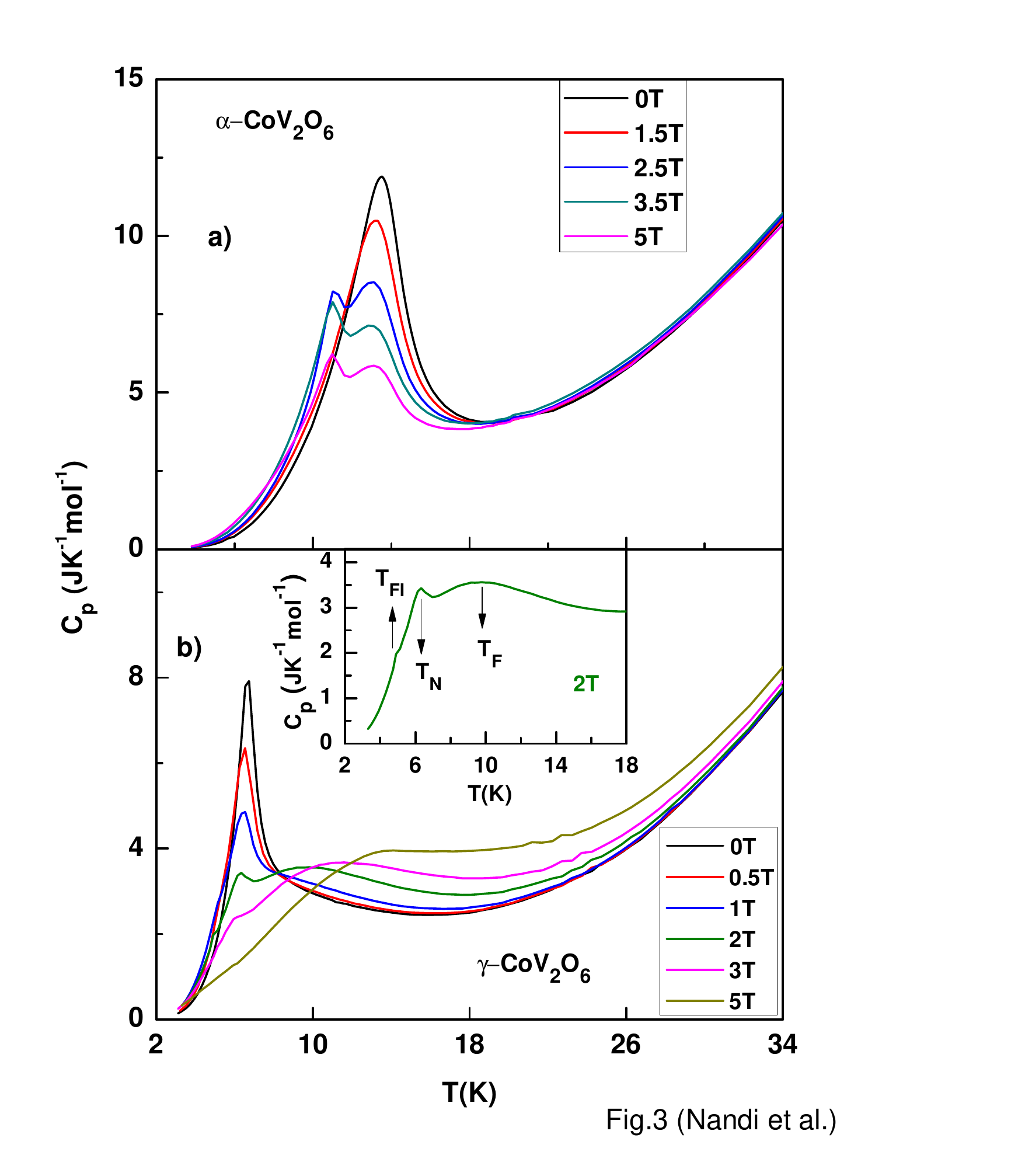}
\caption{(a) and (b) Temperature dependence of specific heat of $\alpha$-CoV$_{2}$O$_{6}$ and $\gamma$-CoV$_{2}$O$_{6}$, are plotted for selected fields, respectively. Inset: Zoomed view of $C_p$($T$) at 2 T of $\gamma$-CoV$_{2}$O$_{6}$, indicating three anomalies around three transition temperatures $T_N$, $T_{FI}$ and $T_F$.}\label{fig3}
\end{figure}

\begin{figure}[h]
\includegraphics[width=0.6\textwidth]{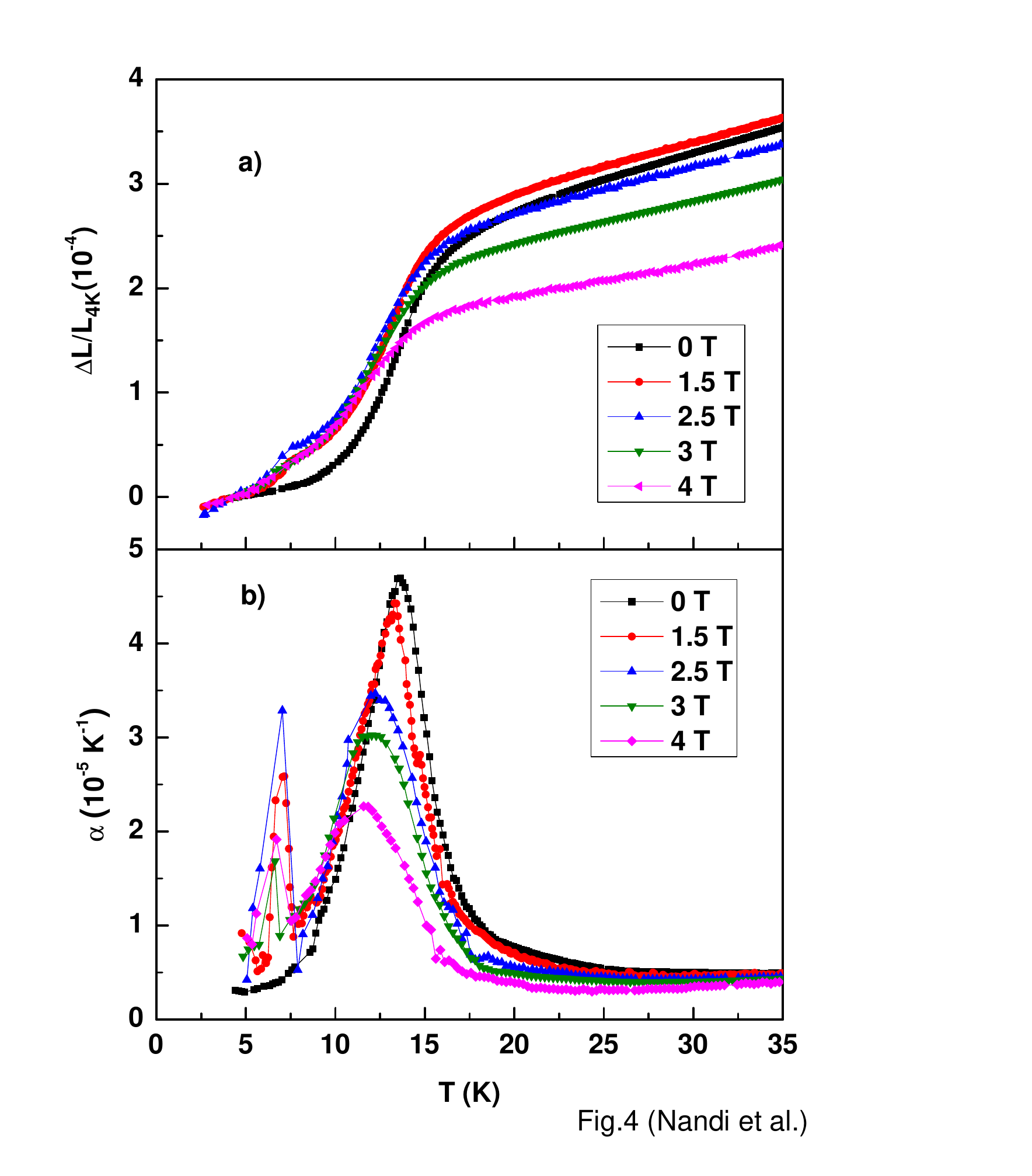}
\caption{(a) The temperature dependence of the relative length change ($\Delta L/L_{4K}$) for $\alpha$-CoV$_2$O$_6$ for different fields. (b) Plots of the thermal expansion coefficient $\alpha$($T$) for different fields.}\label{fig4}
\end{figure}

\begin{figure}[h]
\includegraphics[width=0.6\textwidth]{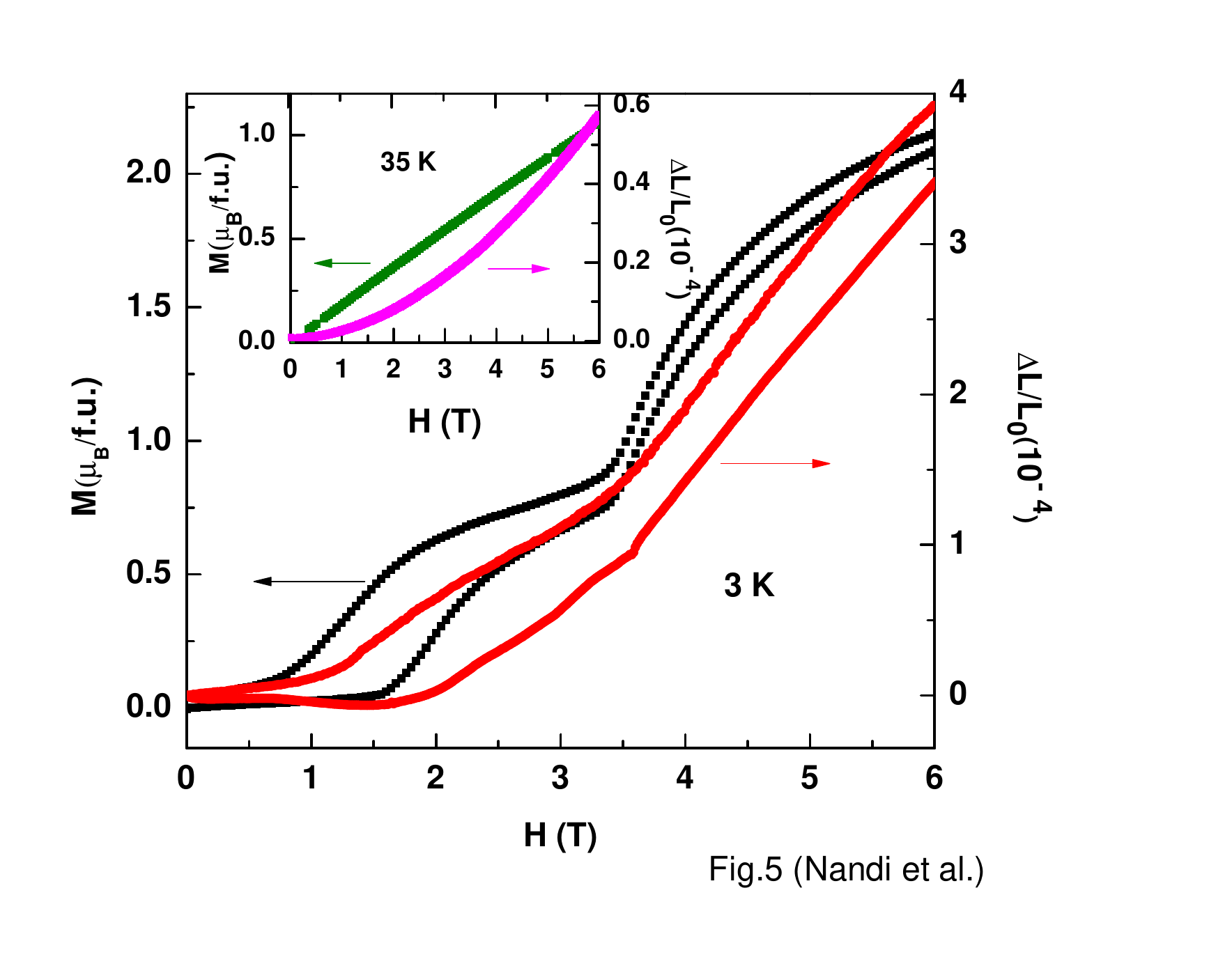}
\caption{Magnetostriction, $\Delta L(H)/L_0$, and magnetization for both increasing and decreasing field at 3 K for $\alpha$-CoV$_2$O$_6$. Inset: the field dependence of magnetization and $\Delta L/L_0$ at 35 K.}\label{fig5}
\end{figure}


\newpage
\begin{thebibliography}{99}
\bibitem{maignan} A. Maignan, V. Hardy, S. Hebert, M. Drillon, M. R. Lees, O. Petrenko, D. M. K. Paul, and D. Khomskii, J. Mater. Chem. 14 (2004) 1231.
\bibitem{maignan1}  A. Maignan, C. Michel, A. C. Masset, C. Martin, and B. Raveau, Eur. Phys. J. B 15 (2000) 657.
\bibitem{koba} H. Kobayashi and T. Haseda, J. Phys. Soc. Jpn. 19 (1964) 765.
\bibitem{cold} R. Coldea, D. A. Tennant, E. M. Wheeler, E. Wawrzynska, D. Prabhakaran, M. Telling, K. Habicht, P. Smeibidl, and K. Kiefer, Science 327 (2010) 177.
\bibitem{he} Z. He, T. Taniyama, K$\hat{y}$omen, and M. Itoh, Phys. Rev. B 72 (2005) 172403.


\bibitem{holl} N. Hollmann, S. Agrestini, Z. Hu, Z. He, M. Schmidt, C.-Y. Kuo, M. Rotter, A. A. Nugroho, V. Sessi, A. Tanaka, N. B. Brookes, and L. H. Tjeng, Phys. Rev. B  89 (2014) 201101(R).
\bibitem{kim} B. Kim, B. H. Kim, K. Kim, H. C. Choi, S. Y. Park, Y. H. Jeong, and B. I. Min, Phys. Rev. B 85 (2012) 220407(R).
\bibitem{lene} M. Lenertz, J. Alaria, D. Stoeffler, S. Colis, and A. Dinia, J. Phys. Chem. C 115 (2011) 17190.

\bibitem{zhe} Z. He, J. Yamaura, Y. Ueda, and W. Cheng, J. Am. Chem. Soc. 131 (2009) 7554.
\bibitem{lener} M. Lenertz, J. Alaria, D. Stoeffler, S. Colis, and A. Dinia, Phys. Rev. B 86 (2012) 214428.
\bibitem{ksingh} K. Singh, A. Maignan, D. Pelloquin, O. Perez, and Ch. Simon, J. Mater. Chem. 22 (2012) 6436.
\bibitem{mark} M. Markkula, A. M. Arevalo-Lopez, and J. P. Attfield, J. Solid State Chem. 192 (2012) 390.
\bibitem{markk} M. Markkula, A. M. Arevalo-Lopez, and J. P. Attfield, Phys. Rev. B 86 (2012) 134401.
\bibitem{nandi} M. Nandi, N. Khan, D. Bhoi, A. Midya and P. Mandal,  J. Phys. Chem. C 118 (2014) 1668.

\bibitem{kimb}  S. A. J. Kimber, H. Mutka, T. Chatterji, T. Hofmann, P. F. Henry, H. N. Bordallo, D. N. Argyriou, and J. P. Attfield, Phys. Rev. B 84 (2011) 104425.
\bibitem{zh} Z. He and M. Itoh,  J. Crystal Growth 388 (2014) 103.
\bibitem{lenert} M. Lenertz, S. Colis, C. Ulhaq-Bouillet and A. Dinia, Appl. Phys. Lett. 102 (2013) 212407.
\bibitem{lenertz}M. Lenertz, A. Dinia, and S. Colis, J. Phys. Chem. C 118 (2014) 13981.
\bibitem{Drees}Y. Drees, S. Agrestini, O. Zaharko, and A. C. Komarek, Cryst. Growth Des.  {\bf 15}, 1168 (2015).

\end{thebibliography}
\end{document}